\documentclass[preprint]{aastex}
\usepackage{natbib}

\shorttitle{HST Imaging of Bright LBG Candidates}
\shortauthors{Bentz, et al.}

\received{}
\accepted{}

\begin{document}

\title{{\it Hubble Space Telescope} Imaging of Bright Lyman-break Galaxy 
Candidates from the Sloan Digital Sky Survey: Not LBGs After All}

\author{ Misty~C.~Bentz\altaffilmark{1,2},
Richard~W.~Pogge\altaffilmark{1},
Patrick~S.~Osmer\altaffilmark{1}}

\altaffiltext{1}{Department of Astronomy, 
		The Ohio State University, 
		140 West 18th Avenue, 
		Columbus, OH 43210; 
		pogge, osmer@astronomy.ohio-state.edu}

\altaffiltext{2}{Present address:
                 Department of Physics and Astronomy,
		 4129 Frederick Reines Hall,
		 University of California, Irvine
		 Irvine, CA 92697;
		 mbentz@uci.edu}

\begin{abstract}

We present deep {\it Hubble Space Telescope} ACS and NICMOS images of
six bright Lyman-break galaxy candidates that were previously
discovered in the Sloan Digital Sky Survey.  We find that five of the
objects are consistent with unresolved point sources.  Although
somewhat atypical of the class, they are most likely LoBAL quasars,
perhaps FeLoBALs.  The sixth object, J1147, has a faint companion
galaxy located $\sim 0.8$\,arcsec to the southwest.  The companion
contributes $\sim 8\%$ of the flux in the observed-frame optical and
infrared.  It is unknown whether this companion is located at the same
redshift as J1147.

\end{abstract}


\section{INTRODUCTION}

The discovery of an apparently luminous (r $\approx 20.5$ mag)
star-forming galaxy at z $\approx 2.5$ by \citet{bentz04a} in the
Early Data Release \citep{stoughton02} of the Sloan Digital Sky Survey
(SDSS, \citealt{york00}) Quasar Catalog \citep{schneider02} was an
unexpected surprise.  Most known Lyman break galaxies (LBGs) have been
discovered in deep images covering very small areas of the sky, and
the brightest LBGs have R$_{AB} \approx 23$ magnitudes, which is much
fainter than the limit of SDSS.  A dedicated search through the SDSS
First Data Release (DR1, \citealt{abazajian03}) produced five
additional objects, as well as the original EDR object
(\citealt{bentz04b}, see Figure 1), with redshifts $2.45 < z < 2.80$
and $19.8 < m_r < 20.5$.  These six objects, by definition, have
similar emission- and absorption-line properties to the LBG composite
spectrum produced from 811 individual objects \citep{shapley03}, but
there are slight differences as well, in that the spectra are redder,
the widths of the lines are greater, and the high ionization lines are
stronger than the lower ionization lines.  While such differences
bring into question the true nature of the underlying source of power
for these objects, the different possibilities are all extremely
important.

The first possibility is that these objects truly are ultraluminous
LBGs, over 4 mag brighter than an ``$L_*$'' LBG.  Using the continuum
luminosity at $\lambda$1500 \AA\ as a proxy for the star formation,
\citet{bentz04b} calculated the star formation rates (SFRs) to be
$\sim 300-1000$ M$_{\odot}$ yr$^{-1}$, {\it without} corrections for
dust.  As these spectra are quite red, the amount of light attenuated
by dust can be expected to be higher than the typical factor of 7
determined by \citet{shapley03} for LBGs, which would indicate
significantly higher SFRs.  The turbulence of such a system may be
able to reproduce the differences observed in the spectra of the
objects compared to the composite spectrum.  A rough estimate of the
luminosity function of these objects is well above a
\citet{schechter76} function extrapolation of the LBG luminosity
function determined by \citet{adelberger00}.  Such objects would be
the most luminous members of their class known, and would be extremely
important to understanding the formation and evolution of star-forming
galaxies at their peak epoch of $z \approx 3$.
Furthermore, because the space density of these objects is higher than
predicted, they could represent a new class of objects or activity.

The second possibility is that these objects are star-forming
galaxies, but they are significantly brightened by gravitational
lensing.  This has already been found to be the case with the galaxy
MS~1512-cB58, which is an ``$L_*$'' LBG magnified by a factor of 30 by
a foreground cluster \citep{seitz98}.  The lensing hypothesis would
account for the space density being higher than predicted for objects
this luminous, but it does not seem to be borne out by the differences
in the spectra of the objects when compared with the LBG composite of
\citet{shapley03}.

Finally, it is possible that these objects are an unusual class of
active galactic nuclei (AGNs) that has not been previously seen.  The
greater line widths in the spectra of the candidates when compared
with the composite spectrum could be evidence for the AGN hypothesis,
and one object has a hint of a broad \ion{C}{3} emission line.  The
main argument against this hypothesis has been the high level of
similarity between the spectra of the objects and the LBG composite
spectrum.

With the SDSS data alone, the exact mechanism behind the source of
power in these objects cannot be determined.  While there is no
evidence for lensing in the SDSS images, and the fields surrounding
the objects seem quite empty, groups or clusters at z $\approx 1$
would be too faint to be detected by SDSS.  Also, the typical point
spread function (PSF) of the SDSS images is $\sim$1.4'', which would
mask any distortions of the source caused by gravitational lensing,
and also effectively confuses the distinction between a point source
or an extended object with a diameter on the scale of the PSF.
Furthermore, the low S/N of the SDSS spectra masks the finer details
in the rest-frame UV spectra and thus prevents a decision on whether
the source of power is dominated by star formation or AGN emission.

In this paper, we present deep {\it HST} follow-up images of these six
candidates discovered by \citet{bentz04b} in the rest-frame UV and
optical.  We apply two-dimensional image decomposition techniques to
determine the morphological classifications of each of the objects and
the relative contributions of the fitted components to the flux of
each object.  We find that five of the objects are consistent with
unresolved point sources (i.e., AGNs), and that the sixth has a faint,
nearby companion that contributes $\sim 8$\% of the rest-frame UV flux
measured through the SDSS fiber.  Finally, we discuss the results of
our analysis in the context of additional follow-up studies that have
meanwhile been carried out on these unusual objects.




\section{OBSERVATIONS}

Each of the six candidate bright LBGs listed in Table 1 were observed
for a total of three orbits with the Hubble Space Telescope ({\it
HST}): two orbits with the Advanced Camera for Surveys Wide-Field
Channel (ACS WFC), and one orbit with the Near-Infrared Camera and
Multi-Object Spectrometer (NICMOS).  With the ACS WFC we acquired deep
$r$-band images with the goal of studying the rest-frame UV morphology
of the objects, and to search for evidence of possible foreground
lensing of the objects.  Deep $H$-band images with the NICMOS NIC2
camera were acquired to try to detect the host galaxy in the
rest-frame optical, as well as to search for any nearby galaxies in
proximity to our primary targets.  We present the details of the
observations for each in the following sections.

\subsection{Advanced Camera for Surveys}

Each of the six targets was observed with two full {\it HST} orbits
using the ACS WFC through the F625W filter (similar to the SDSS $r$
filter).  The F625W filter was chosen specifically to take advantage
of the maximum throughput of the optics at 6300~\AA\ to get deep
images of the fields in the rest-frame UV and any foreground groups or
clusters.  The higher sensitivity of the WFC was deemed more important
in this case than the resolution that would be afforded by the High
Resolution Channel on ACS.  Table~1 lists the total exposure time for
each object, which was broken up into five individual exposures of
$\sim 14$ minutes each.  Between each of the five exposures, the
telescope was dithered in a line pattern, with a typical offset of
3\,arcsec per dither, to assist in the rejection of cosmic rays and
bad pixels.

The standard {\it HST} reduction and calibration pipeline did a fine
job of processing and combining the individual images, which are
presented in the left hand panels of Figures~$1-6$, rotated and scaled
to match the field of view of NICMOS.

\subsection{Near Infrared Camera and Multi-Object Spectrometer}

Each of the six objects was also observed for one full orbit with the
NIC2 camera on NICMOS through the F160W (broad-band $H$) filter.  The
observations were scheduled to avoid the cosmic ray persistence
problem caused by the South Atlantic Anomaly.\footnote{See ISR
NICMOS-98-001.}  The F160W filter was chosen to take advantage of the
fact that the rest-frame visible spectrum is redshifted into the
infrared and would be the best way to study any visible hosts and
companions of our targets.

The observations were acquired in MULTIACCUM mode, utilizing the {\it
step64} sample sequence, which has a series of rapid, non-destructive
reads up to 64\,s, and then reads out in steps of 64\,s.  Such a
sample sequence gives a high dynamic range in the resulting image,
which allows for the study of both very faint and very bright objects
in the same field.  The total exposure time for each target is listed
in Table~1, and is typically composed of five separate exposures of
511\,s.  For J1147 we were able to acquire two additional exposure of
243\,s, and for J1340, a total of seven exposures of 384\,s each were
acquired.  Between each individual exposure, the spacecraft was
dithered in a spiral pattern, with a typical offset of 1\,arcsec.

The images were reduced and combined in the usual way by the {\it HST}
calibration pipeline, and are presented in the right-hand panels of
Figures~$1-6$.

\section{IMAGE DECOMPOSITIONS}

At first glance, all six of the objects appear to be unresolved in
both the rest-frame UV and optical, although there does appear to be a
faint companion to J1147 that is slightly offset from the PSF of the
unresolved target.  Within a radius of $\sim 20$\,arcsec, the fields
of these objects appear rather unremarkable, although there may be a
slight overdensity in the field of J1432.

To search for any faint host galaxy contribution that may be hiding
underneath the bright PSF, we employed the two-dimensional image
decomposition program Galfit \citep{peng02}, which fits analytic
functions to an image, convolved with a user-supplied PSF model.
Model PSFs were created using the TinyTim software package, which
models the optics of {\it HST} plus the specifics of the camera and
filter system \citep{krist93}.  For the ACS images, a model PSF was
created for the location of the target in each individual exposure,
and the model PSFs were then drizzled together and corrected for the
distortion in the optics of the camera using the same methods as were
applied to the data.

As each of the targets is very compact and the fields are relatively
empty, we initially fit the images of each target with a central PSF
and a sky background that was allowed to tilt.  This simple model
gives a reasonable fit to both the ACS and NICMOS image of each
target, except for J1147, which clearly has a faint companion offset a
few pixels ($\sim 0.8$\,arcsec) to the southwest of center of the PSF.
However, in all cases, a slightly better fit was achieved with the
addition of a compact \citep{sersic68} profile, which has the form
\begin{equation}
\Sigma (r) = \Sigma_e \exp^{-\kappa [(r/r_e)^{1/n}-1]} 
\end{equation}
where $r_e$ is the effective radius of the component, $\Sigma_e$ is
the surface brightness at $r_e$, $n$ is the power-law index, and
$\kappa$ is coupled to $n$ such that half of the total flux is within
$r_e$. Two special cases of the S\'{e}rsic function are the
exponential profile ($n=1$), often used in modeling galactic disks,
and the \citet{devaucouleurs48} profile ($n=4$), historically used for
modeling galactic bulges. Table~2 lists the integrated magnitudes and
best-fit parameters of the PSF and S\'{e}rsic fits to each of the
targets in the ACS and NICMOS images.  Figure~7 compares the residuals
after subtracting a single PSF component from each image to the
residuals after subtracting a PSF component plus a S\'{e}rsic
component.

In addition, we also fit any field objects that were visible in both
the ACS and NICMOS images of each source.  The NIC2 field of view is
much smaller than that of the WFC, therefore, only objects that were
$\lesssim 15$\,arcsec from each target were included.  Table~3 lists
the fit parameters determined for the field objects.  The models and
the fit residuals are shown in the bottom panels of Figures~1-6.

\section{DISCUSSION}

While the fits to the images were statistically improved by the
addition of a S\'{e}rsic component to the PSF model produced by
TinyTim, it is unlikely that these S\'{e}rsic components represent a
separate flux component, and are more likely due to PSF mismatches
between the data and the models.  Indeed, for J0243, J1444, and J1553,
the extra flux component attributed to the S\'{e}rsic profile lies on
the right-hand side of the PSF in the unrotated ACS image.  This is
also true of the NICMOS images presented in Figure~7, where the PSF
mismatch is most apparent along the top edge of the PSF model for not
only these three, but all six objects.

Rather, it appears that the only object with a significant
contribution of flux from a galaxy component is J1147, which has a
faint galaxy companion centered $\sim 0.8$\,arcsec SW of the PSF.
Although it is obviously visible in the images, this additional
component only contributes $\sim 8\%$ of the flux to both the
observed-frame optical and infrared.  While it does have a similar
$r-H$ color to J1147, $-0.09$ and $-0.13$, respectively, with such
limited information it is impossible to know whether this object is at
the same redshift as J1147.

In addition, there is also no evidence to support the lensing
hypothesis for these objects.  The fields around them are fairly empty
for the most part.  J1432 has a modest overdensity of objects, but
there are no lensing artifacts, such as arcs or rings, present in
either the ACS or NICMOS images of this object.

From the fact that all of these objects (with the possible exception
of J1147) are unresolved point sources in both deep optical and
infrared images at the resolution of {\it HST}, it appears that these
objects are unusual AGNs.  Indeed, this is consistent with the
findings of \citet{appenzeller05}, who aquired an echelle spectrum of
J1553 on the VLT and found that it was a broad absorption line quasar
mimicking a Lyman-break galaxy through the combination of absorption
lines with relatively moderate widths and unfortunately located metal
lines from an intervening system.  They identify J1553 as most likely
being a member of the rare iron low-ionization broad absorption line
(FeLoBAL) quasar class.  But even for this class of objects, J1553 is
still an oddity.  Furthermore, near-infrared spectroscopy by
\citet{ivison05} shows that each of the objects has a very broad
H$\alpha$ emission line, which is a clear sign of AGN activity.
Finally, \citeauthor{ivison05} were unable to detect any of the six
objects in the submillimeter, indicating that they cannot possibly
have the high SFRs that would naturally accompany such luminous
objects if they were bona-fide LBGs.

\section{SUMMARY}

We have presented high-resolution {\it HST} follow-up images of the
six bright Lyman-break galaxy candidates that were previously
identified in the Sloan Digital Sky Survey.  We find that, with the
exception of J1147, they are all consistent with unresolved point
sources in both the rest-frame UV and optical.  They are most likely
members of the LoBAL quasar class, and possibly the even rarer FeLoBAL
class, although their spectral properties are atypical of both classes
of objects.

In the case of J1147, an additional flux component has been
identified, located a mere \mbox{$\sim 0.8$\,arcsec} SW of the PSF,
but this component is only contributing $8\%$ of the flux.  Further
study will be necessary to determine whether this object is located at
the same redshift as J1147.

\acknowledgements
This work is based on observations with the NASA/ESA {\it Hubble Space
Telescope}.  We are grateful for support of this work through grant
GO-10181 from the Space Telescope Science Institute, which is operated by
the Association of Universities for Research in Astronomy, Inc., under
NASA contract NAS5-26555.

\clearpage


\clearpage

\begin{figure}
\plotone{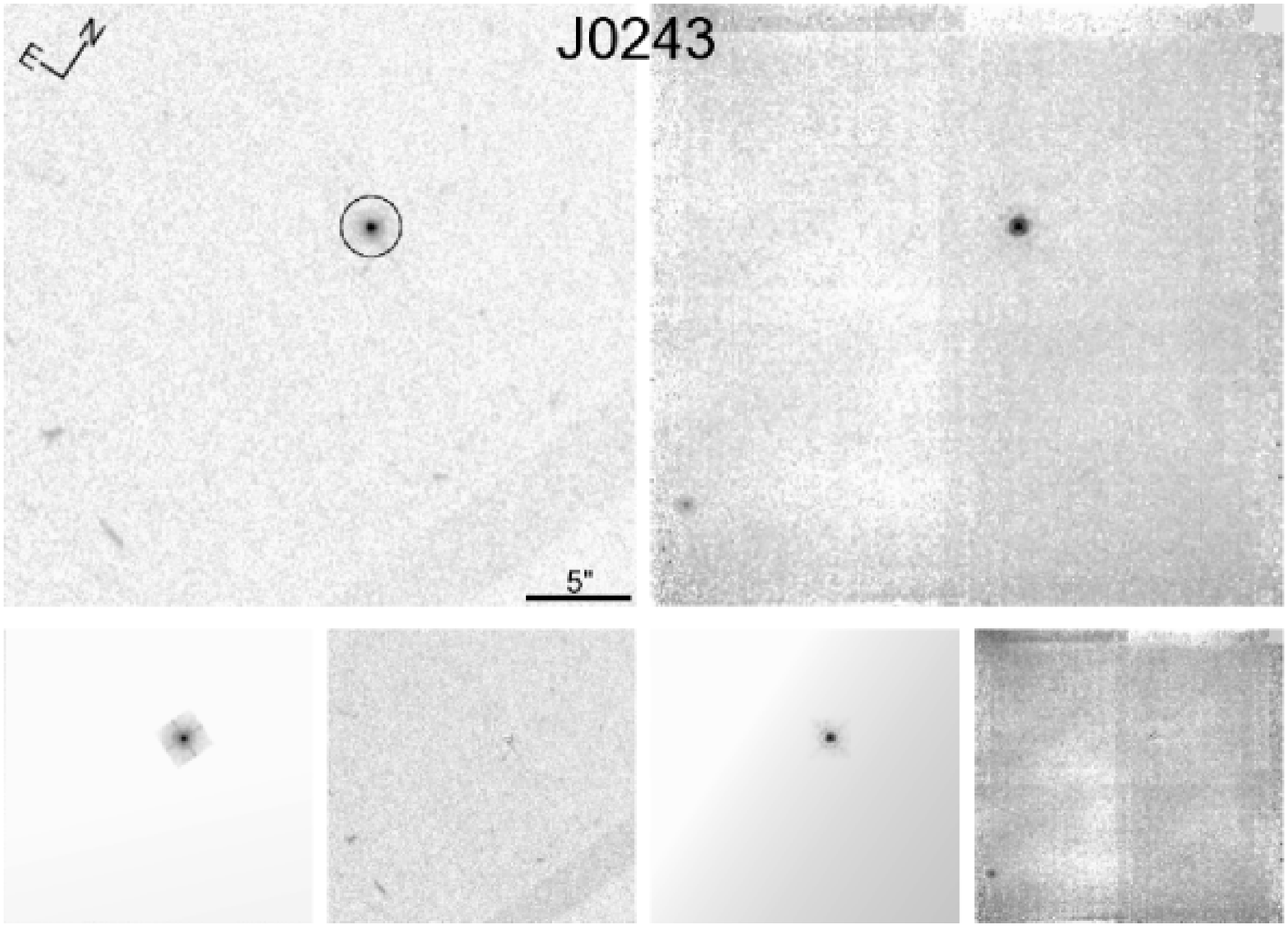}
\caption{ACS imaging of J0243 through the F625W filter (top left
panel) and NICMOS imaging of J0243 through the F160W filter (top right
panel).  The circle around the object in the ACS image shows the size
of the SDSS fiber on the sky.  The ACS images have been scaled and
oriented to match the NICMOS images.  The bottom panels show, from
left to right: Galfit model of the ACS image and any field objects
that appear in both the optical and infrared, Galfit residuals after
subtracting the model from the ACS image, Galfit model of the NICMOS
image of J0243 and any shared field objects, and Galfit residuals
after subtracting the model from the NICMOS image.}
\end{figure}

\begin{figure}
\plotone{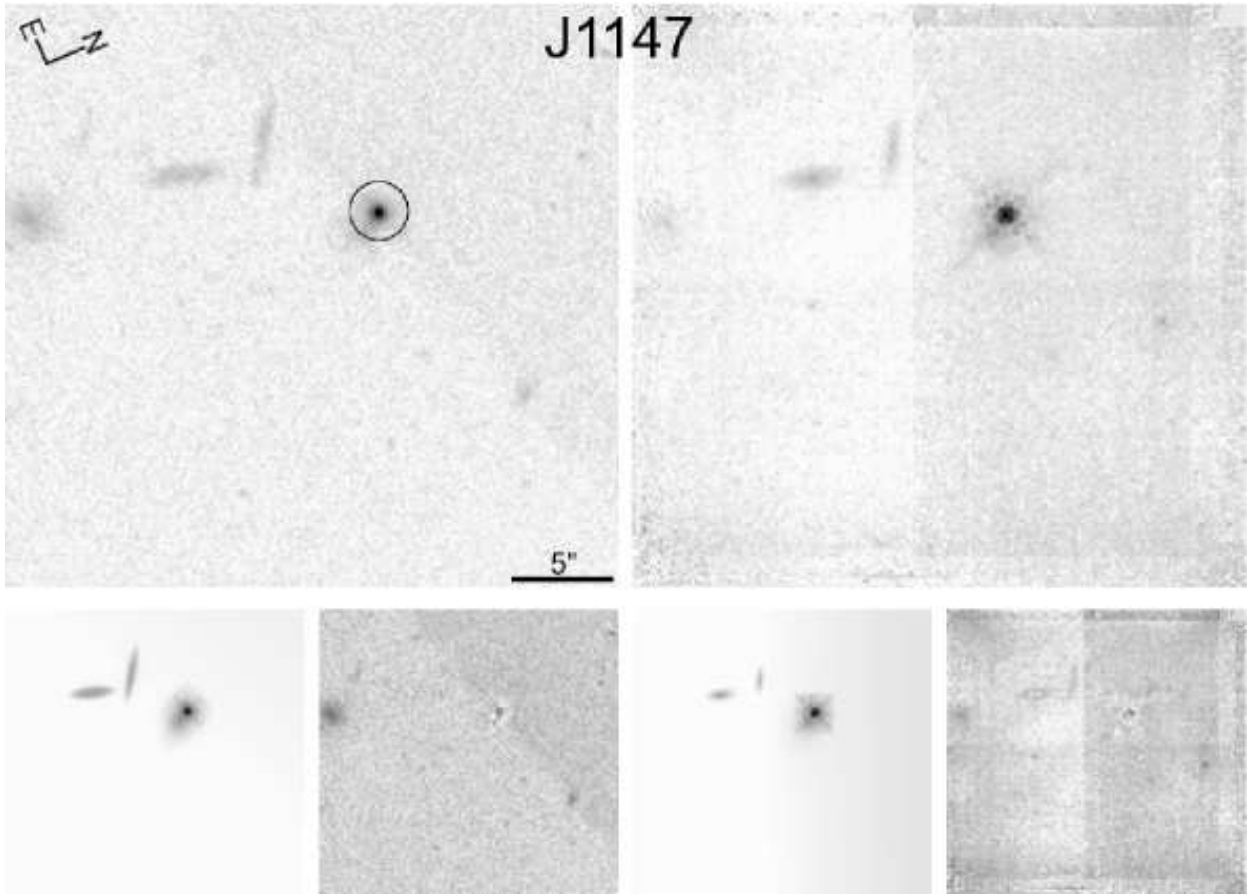}
\caption{Same as Figure 1, but for J1147.  Note the additional flux
component located to the southwest of the PSF.}
\end{figure}

\begin{figure}
\plotone{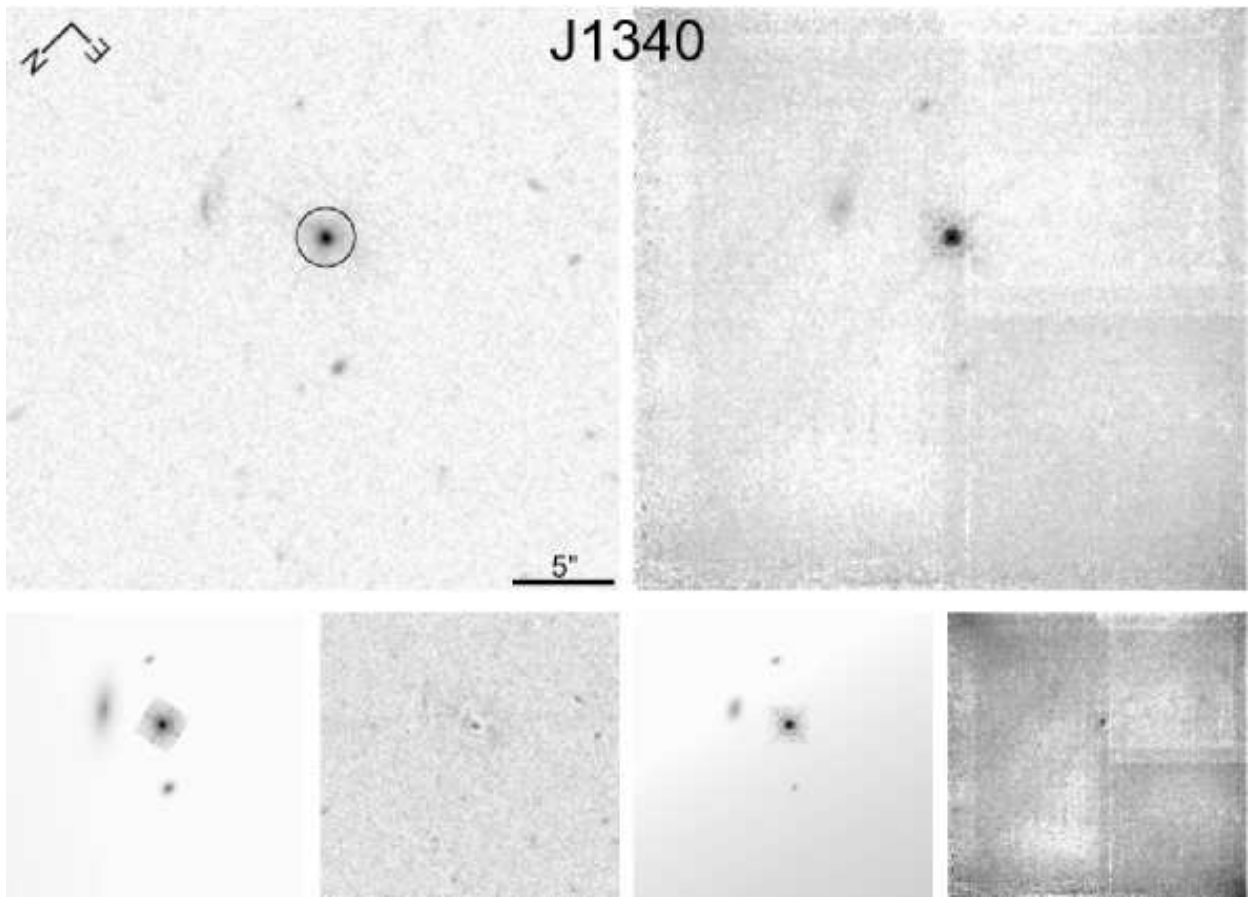}
\caption{Same as Figure 1, but for J1340.}
\end{figure}

\begin{figure}
\plotone{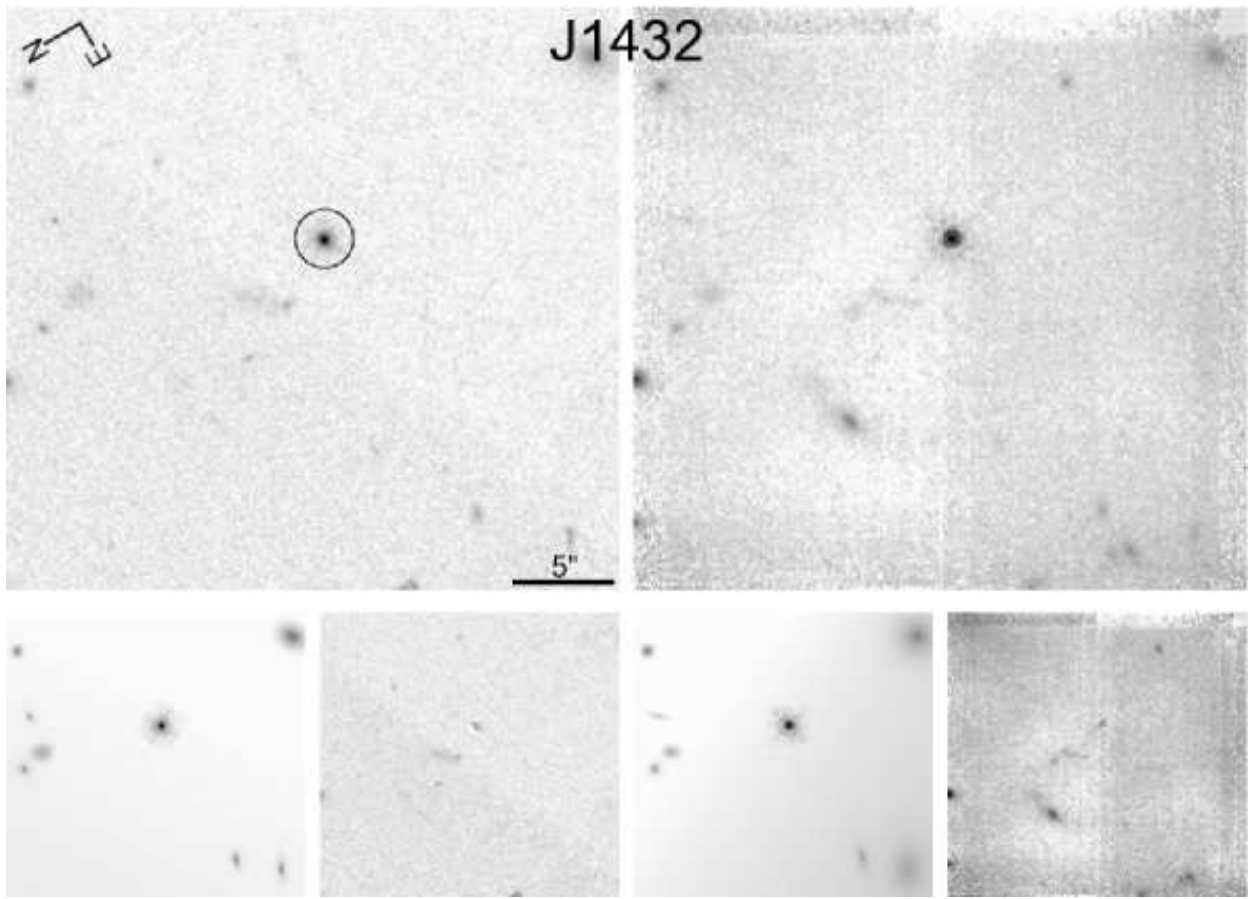}
\caption{Same as Figure 1, but for J1432.}
\end{figure}

\begin{figure}
\plotone{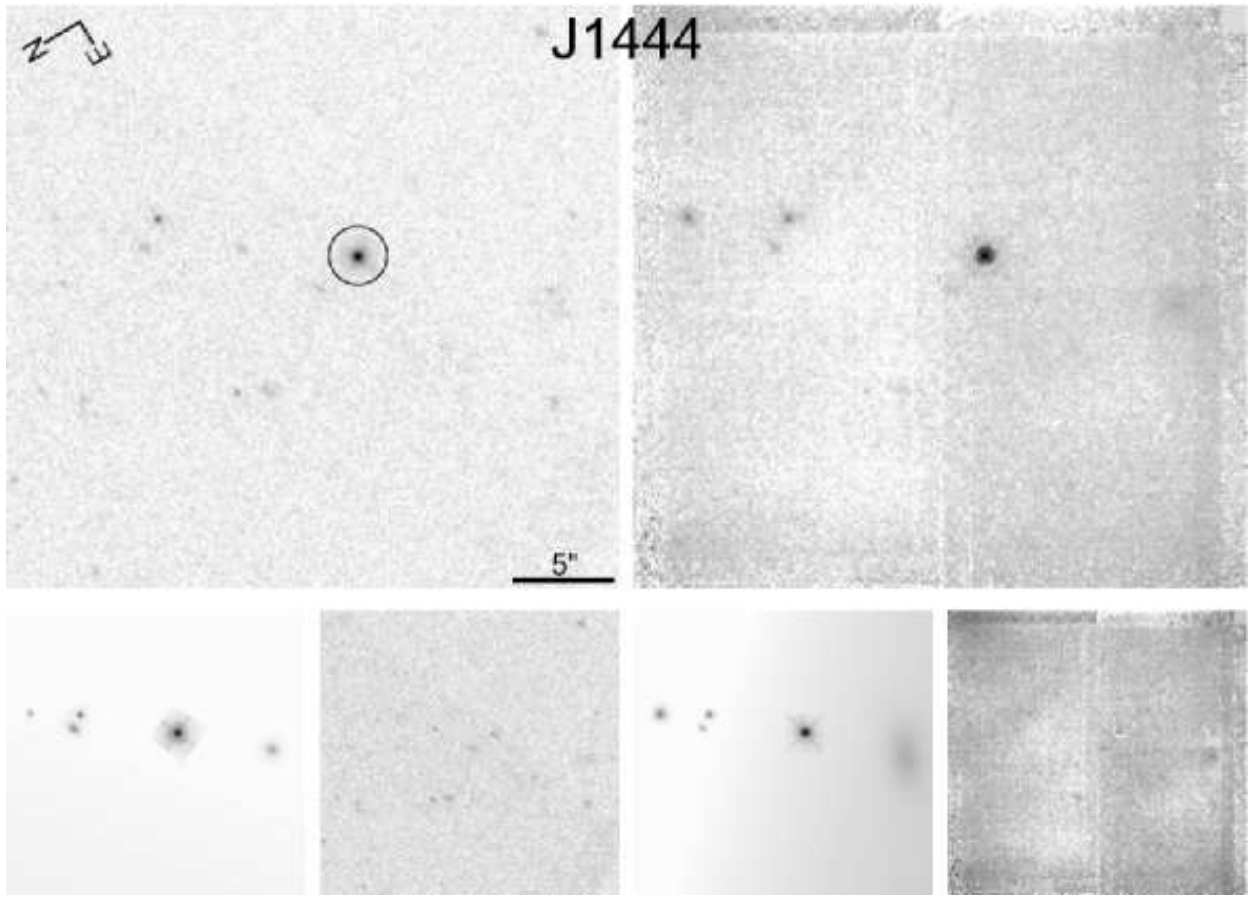}
\caption{Same as Figure 1, but for J1444.}
\end{figure}

\begin{figure}
\plotone{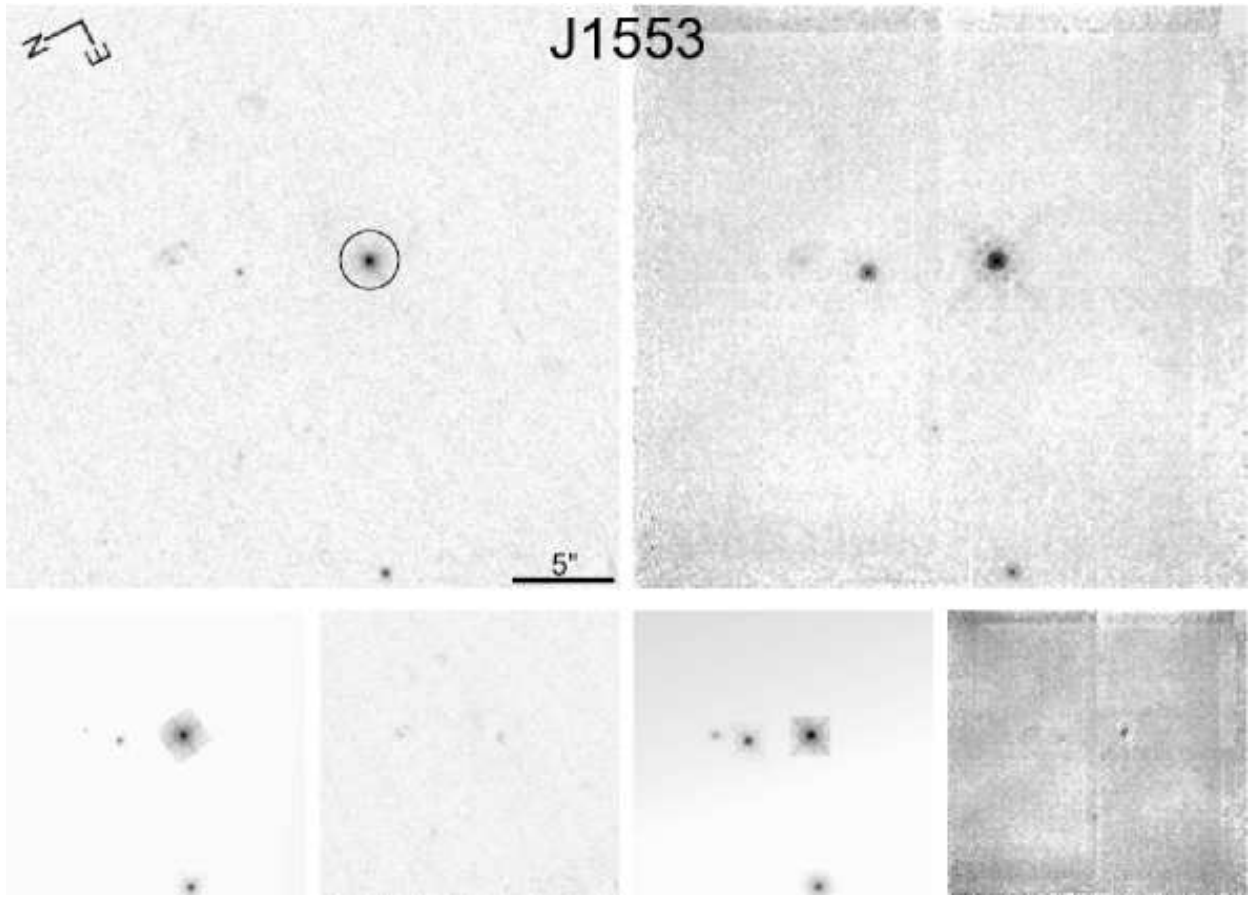}
\caption{Same as Figure 1, but for J1553.}
\end{figure}

\begin{figure}
\epsscale{0.75}
\plotone{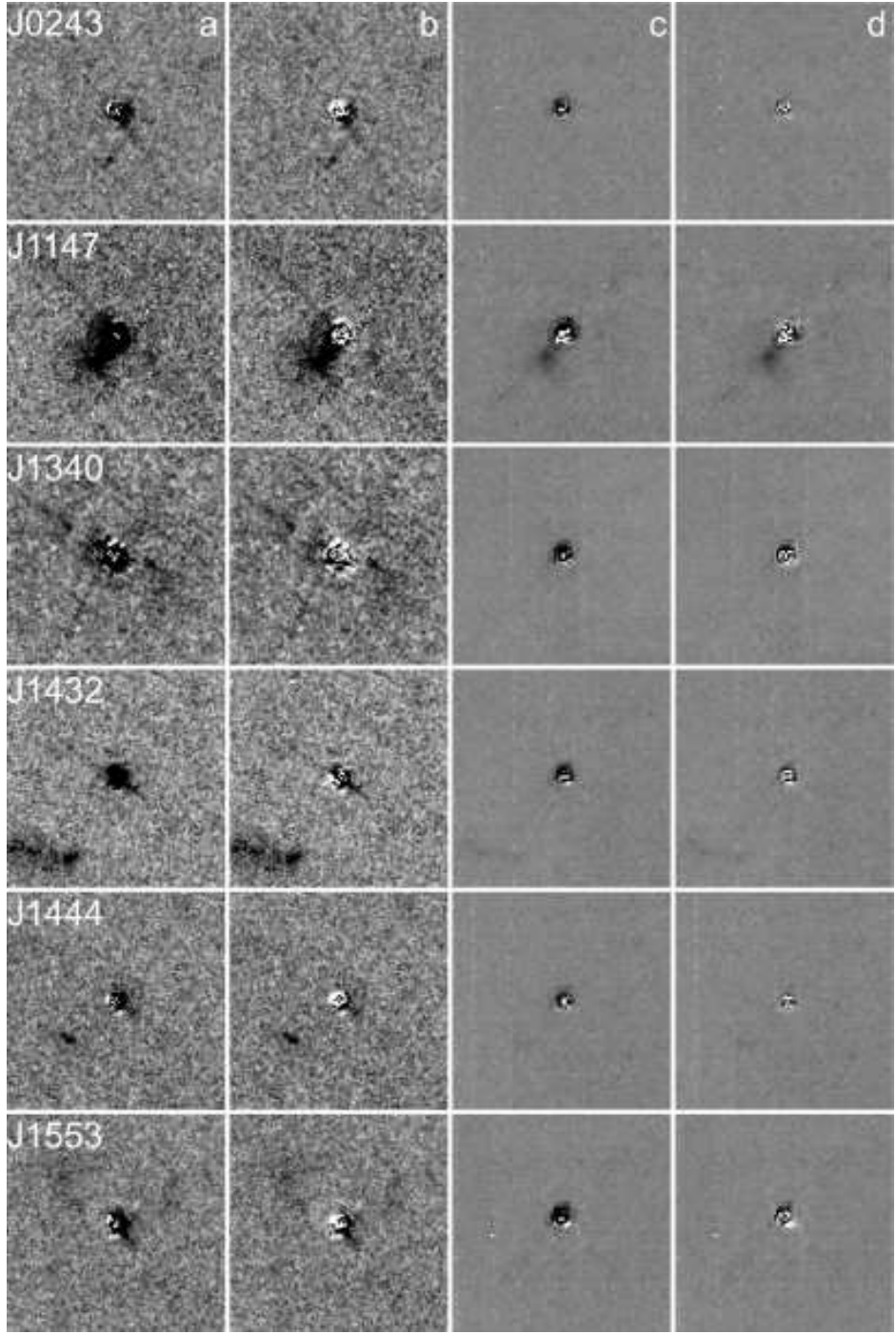}
\caption{Details of the PSF subtraction for each object.  Each panel
is $6'' \times 6''$, and each object is shown at the same orientation
as it was previously presented.  All the ACS images have the same
color scale, and the NICMOS images all have the same color scale.  For
each object, the panels are as follows: {\it (a)} ACS F625W image,
with the best-fit PSF subtracted; {\it (b)} ACS F625W image with the
best-fit PSF + S\'{e}rsic combination subtracted; {\it (c)} NIC2 F160W
image with the best-fit PSF subtracted; {\it (d)} NIC2 F160W image
with the best-fit PSF + S\'{e}rsic combination subtracted.}
\end{figure}  

\clearpage

\begin{deluxetable}{lcccccc}
\tablecolumns{7}
\tablewidth{0pt}
\tablecaption{Observation Log}
\tablehead{
\colhead{} &
\colhead{} &
\multicolumn{2}{c}{ACS WFC F625W} &
\colhead{} &
\multicolumn{2}{c}{NIC2 F160W}\\ \cline{3-4} \cline{6-7}
\colhead{Object} &
\colhead{$z$} &
\colhead{Date Observed} &
\colhead{Exp. Time} &
\colhead{} &
\colhead{Date Observed} &
\colhead{Exp. Time} \\
\colhead{(SDSS J)} &
\colhead{} &
\colhead{(yyyy-mm-dd)} &
\colhead{(s)} &
\colhead{} &
\colhead{(yyyy-mm-dd)} &
\colhead{(s)}}

\startdata

024343.77$-$082109.9  & 2.590  & 2005-01-11 & 4150.0 & & 2004-08-12 & 2559.7 \\
114756.00$-$025023.5  & 2.556  & 2004-11-20 & 4150.0 & & 2004-11-19 & 3046.4 \\
134026.44+634433.2    & 2.786  & 2004-07-12 & 4550.0 & & 2004-07-24 & 2687.7 \\
143224.55$-$000116.4  & 2.472  & 2005-12-27 & 4000.0 & & 2004-08-12 & 2559.7 \\
144424.55+013457.0    & 2.670  & 2005-01-10 & 4150.0 & & 2004-08-12 & 2559.7 \\
155359.96+005641.3    & 2.635  & 2006-03-09 & 4150.0 & & 2004-08-12 & 2559.7 \\

\enddata
\end{deluxetable}

\begin{deluxetable}{lcccccccccccc}
\tablecolumns{13}
\tablewidth{0pt}
\tabletypesize{\small}
\tablecaption{Fit Parameters for LBG Candidates}
\tablehead{
\colhead{} &
\colhead{PSF} &
\multicolumn{5}{c}{Sersic component} &
\colhead{} &
\multicolumn{5}{c}{Additional component}  \\ \cline{3-7} \cline{9-13}
\colhead{Object} &
\colhead{$m$\tablenotemark{a}} &
\colhead{$m$} &
\colhead{$\Delta$ d\tablenotemark{b}} &
\colhead{$r_{\rm e}$} &
\colhead{$n$} &
\colhead{$b/a$\tablenotemark{c}} &
\colhead{} &
\colhead{$m$} &
\colhead{$\Delta$ d} &
\colhead{$r_{\rm e}$} &
\colhead{$n$} &
\colhead{$b/a$} \\
\colhead{} &
\colhead{(stmag)} &
\colhead{(stmag)} &
\colhead{(arcsec)} &
\colhead{(kpc)} &
\colhead{} &
\colhead{} &
\colhead{} &
\colhead{(stmag)} &
\colhead{(arcsec)} &
\colhead{(kpc)} &
\colhead{} &
\colhead{}}
\startdata
\multicolumn{13}{c}{ACS WFC F625W} \\
\hline

J0243 & 21.88 & 21.19 & 0.04 & 0.26 & 1.0  & 0.12 & & & & & & \\
J1147 & 21.37 & 20.92 & 0.06 & 0.26 & 4.5  & 0.79 & & 23.10 & 0.78 & 6.47 & 0.96 & 0.64 \\
J1340 & 20.72 & 20.25 & 0.04 & 0.12 & 4.8  & 0.65 & & & & & & \\ 
J1432 & 22.44 & 20.31 & 0.05 & 0.01 & 14.2 & 0.49 & & & & & & \\ 
J1444 & 21.44 & 22.16 & 0.04 & 0.39 & 1.0  & 0.57 & & & & & & \\
J1553 & 22.11 & 20.83 & 0.03 & 0.31 & 2.8  & 0.48 & & & & & & \\

\hline
\multicolumn{13}{c}{NIC2 F160W}\\
\hline

J0243 & 22.69 & 23.03 & 0.04 & 0.40 & 1.0  & 0.85 & & & & & & \\  
J1147 & 20.59 & 23.38 & 0.17 & 1.71 & 1.0  & 0.24 & & 23.19 & 0.77 & 6.86 & 1.9 & 0.71 \\
J1340 & 22.06 & 21.68 & 0.03 & 0.87 & 0.01 & 0.69 & & & & & & \\
J1432 & 21.88 & 23.31 & 0.08 & 0.68 & 3.2  & 0.73 & & & & & & \\
J1444 & 22.23 & 24.08 & 0.06 & 1.02 & 1.0  & 0.75 & & & & & & \\
J1553 & 21.66 & 22.19 & 0.09 & 0.51 & 1.0  & 0.68 & & & & & & \\

\enddata

\tablenotetext{a}{Magnitudes are presented in the STMAG system, which
is defined such that Vega has a constant flux per unit wavelength and
has the form $m = -2.5 \log f_{\lambda} - 21.10$.}

\tablenotetext{b}{Distance from the center of the component to the
center of the PSF.}

\tablenotetext{c}{Ratio of the semi-major axis to the semi-minor axis
for the component.}

\end{deluxetable}

\begin{deluxetable}{lcccccccccccc}
\tablecolumns{13}
\tablewidth{0pt}
\tabletypesize{\footnotesize}
\tablecaption{Fit Parameters for Field Objects}
\tablehead{
\colhead{} &
\colhead{} &
\multicolumn{5}{c}{ACS HRC F625W} &
\colhead{} &
\multicolumn{5}{c}{NIC2 F160W} \\ \cline{3-7} \cline{9-13}
\colhead{Field} &
\colhead{function} &
\colhead{$\Delta$ d} &
\colhead{$m_{\rm F625W}$} &
\colhead{$r_{\rm e}$} &
\colhead{$n$} &
\colhead{$b/a$} &
\colhead{} &
\colhead{$\Delta$ d} &
\colhead{$m_{\rm F160W}$} &
\colhead{$r_{\rm e}$} &
\colhead{$n$} &
\colhead{$b/a$} \\
\colhead{} &
\colhead{} &
\colhead{(arcsec)} &
\colhead{(mag)} &
\colhead{(arcsec)} &
\colhead{} &
\colhead{} &
\colhead{} &
\colhead{(arcsec)} &
\colhead{(mag)} &
\colhead{(arcsec)} &
\colhead{} &
\colhead{}}
\startdata

J1147 & S\'{e}rsic & 4.71    & 23.86   & 1.04    & 0.36    & 0.16    &&  4.51  & 25.27 & 0.50    & 0.37    & 0.13 \\
      & S\'{e}rsic & 6.80    & 23.53   & 0.88    & 0.25    & 0.26    &&  6.76  & 24.28 & 0.29    & 1.0     & 0.32 \\
J1340 & S\'{e}rsic & 4.28    & 23.51   & 0.88    & 1.8     & 0.37    &&  4.00  & 24.33 & 0.45    & 0.47    & 0.54 \\
      & S\'{e}rsic & 4.47    & 24.17   & 0.14    & 1.2     & 0.67    &&  4.45  & 26.35 & 0.12    & 0.25    & 0.31 \\
      & S\'{e}rsic & 4.70    & 25.94   & 0.11    & 1.9     & 0.44    &&  4.65  & 25.50 & 2.0     & 1.6     & 0.35 \\
J1432 & S\'{e}rsic & 8.57    & 24.29   & 0.43    & 0.45    & 0.82    &&  8.53  & 25.20 & 0.36    & 0.45    & 0.70 \\
      & S\'{e}rsic & 9.24    & 24.82   & 0.37    & 20      & 0.56    &&  9.28  & 25.76 & 0.68    & 0.97    & 0.18 \\
      & S\'{e}rsic & 10.1    & 24.59   & 0.22    & 5.0     & 0.81    &&  9.94  & 25.07 & 0.26    & 3.5     & 0.84 \\
      & S\'{e}rsic & 10.7    & 24.69   & 0.28    & 1.1     & 0.50    &&  10.7  & 24.32 & 1.1     & 2.9     & 0.33 \\
      & S\'{e}rsic & 11.2    & 22.52   & 0.44    & 1.2     & 0.76    &&  11.0  & 22.15 & 1.2     & 1.7     & 0.95 \\
      & S\'{e}rsic & 11.4    & 24.42   & 0.16    & 1.1     & 0.81    &&  11.3  & 24.20 & 0.22    & 3.3     & 0.94 \\
      & S\'{e}rsic & 13.1    & 24.49   & 0.34    & 2.9     & 0.32    &&  13.2  & 22.76 & 1.5     & 0.94    & 0.64 \\
J1444 & S\'{e}rsic & 6.73    & 25.58   & 0.33    & 2.3     & 0.84    &&  7.29  & 22.84 & 1.9     & 0.73    & 0.54 \\
      & S\'{e}rsic & 6.96    & 24.60   & 0.040   & 0.72    & 0.84    &&  6.87  & 25.41 & 0.068   & 2.4     & 0.90 \\
      & S\'{e}rsic & 7.31    & 25.64   & 0.22    & 0.93    & 0.11    &&  7.25  & 26.61 & 0.17    & 0.32    & 0.27 \\
      & S\'{e}rsic & 10.4    & 27.00   & 0.20    & 0.020   & 0.57    &&  10.3  & 23.65 & 1.4     & 8.1     & 0.95 \\
J1553 & PSF        & 4.45    & 25.22   & \nodata & \nodata & \nodata &&  4.41  & 23.43 & \nodata & \nodata & \nodata \\
      & S\'{e}rsic & \nodata & \nodata & \nodata & \nodata & \nodata &&  4.46  & 24.83 & 0.11    & 0.020   & 0.34 \\
      & S\'{e}rsic & 6.85    & 25.22   & 0.00050 & 0.39    & 0.10    &&  6.69  & 26.31 & 0.17    & 1.1     & 0.61 \\
      & S\'{e}rsic & 10.7    & 23.35   & 0.039   & 0.042   & 0.66    &&  10.6  & 23.48 & 2.4     & 6.3     & 0.90 \\
      & S\'{e}rsic & \nodata & \nodata & \nodata & \nodata & \nodata &&  10.6  & 24.44 & 0.013   & 0.82    & 0.90 \\

\enddata
\end{deluxetable}

\end{document}